\documentstyle[12pt,aaspp4,epsfig]{article}
\def\deg{\ifmmode^\circ\else$^\circ$\fi}

\def\mic{~$\mu$m}

\def\eg{{\it e.g.}}

\def\et{{et al.~}}

\def\h0{H$_0$}
\def\q0{q$_0$}

\def\arcs{\ifmmode {''}\else $''$\fi}
\def\arcm{\ifmmode {'}\else $'$\fi}
\def\parcs{\sa=.07em \sb=.03em
     \ifmmode $\rlap{.}$^{\scriptscriptstyle\prime\kern -\sb\prime}$\kern
-\sa$
     \else \rlap{.}$^{\scriptscriptstyle\prime\kern -\sb\prime}$\kern
-\sa\fi}
\def\parcm{\sa=.08em \sb=.03em
     \ifmmode $\rlap{.}\kern\sa$^{\scriptscriptstyle\prime}$\kern-\sb$
     \else \rlap{.}\kern\sa$^{\scriptscriptstyle\prime}$\kern-\sb\fi}

\def\spose#1{\hbox to 0pt{#1\hss}}
\def\simlt{\mathrel{\spose{\lower 3pt\hbox{$\mathchar''218$}}
     \raise 2.0pt\hbox{$\mathchar''13C$}}}
\def\simgt{\mathrel{\spose{\lower 3pt\hbox{$\mathchar''218$}}
     \raise 2.0pt\hbox{$\mathchar''13E$}}}
\def\lsim{\rlap{$<$}{\lower 1.0ex\hbox{$\sim$}}}
\def\gsim{\rlap{$>$}{\lower 1.0ex\hbox{$\sim$}}}

\begin{document}
\slugcomment{Draft Version 0.0:by \today}
\pagestyle{myheadings}
\markboth{DRAFT version 0.0: \today}{DRAFT version 0.0: \today}

\title{J-BAND INFRARED SPECTROSCOPY OF A SAMPLE OF BROWN DWARFS
USING NIRSPEC ON KECK II
\footnote{
Data presented herein were obtained at the W.M. Keck Observatory, 
which is operated
as a scientific partnership among the California Institute of Technology, 
the University of California and the National Aeronautics and Space 
Administration.  The Observatory was made possible by the
generous financial support of the W.M. Keck Foundation.
}}

\author{{Ian S. McLean\altaffilmark{2},}\\
Mavourneen K. Wilcox\altaffilmark{2},
E. E. Becklin\altaffilmark{2}, 
Donald F. Figer\altaffilmark{3},
Andrea M. Gilbert\altaffilmark{4},
James R. Graham\altaffilmark{4},
James E. Larkin\altaffilmark{2},
N. A. Levenson\altaffilmark{5}, 
Harry I. Teplitz\altaffilmark{6,}\altaffilmark{7},
J. Davy Kirkpatrick\altaffilmark{8}.
}

\altaffiltext{2}{Department of Physics and Astronomy, 
University of California, 
%Los Angeles, 
%Division of Astronomy, 
                 Los Angeles, CA, 90095-1562 }
\altaffiltext{3}{Space Telescope Science Institute, 
                  3700 San Martin Dr., Baltimore, MD 21218 }
\altaffiltext{4}{
Department of Astronomy,  
University of California, Berkeley,
601 Campbell Hall,
Berkeley, CA, 94720-3411}
\altaffiltext{5}{
Johns Hopkins University,Department of Physics and Astronomy, Baltimore,
MD  21218}
\altaffiltext{6}{Laboratory for Astronomy and Solar Physics, Code 681,
Goddard
Space Flight Center, Greenbelt MD 20771}
\altaffiltext{7}{NOAO Research Associate}
\altaffiltext{8}{Infrared Processing and Analysis Center, M/S 100-22,
California Institute of
Technology, Pasadena, CA 91125
}

\begin{abstract}
Near-infrared spectroscopic observations of a sample of very cool,
low-mass objects are
presented with higher spectral resolution than in any previous studies.
Six of the objects 
are L-dwarfs, ranging in spectral class from L2 to L8/9, and the seventh
is a methane or T-dwarf. 
These new observations were obtained during commissioning of NIRSPEC, the
first
high-resolution near-infrared cryogenic spectrograph for the Keck II
10-meter telescope on
Mauna Kea, Hawaii. Spectra with a resolving power R$\approx$2500 from
1.135 to
1.360\mic~(approximately J-band) are presented for each source. At this
resolution, a rich
spectral structure is revealed, much of which is due to blending of
unresolved molecular
transitions. Strong lines due to neutral potassium (K I), and bands due to
iron hydride (FeH) and
steam (H$_{2}$O) change significantly throughout the L sequence. Iron
hydride disappears
between L5 and L8, the steam bands deepen and the K I lines gradually
become weaker but
wider due to pressure broadening. An unidentified feature occurs at 1.22\mic~ 
which has a temperature dependence like FeH but has no counterpart in the available
FeH opacity data. 
Because these objects are 3-6 magnitudes
brighter in the near-infrared compared to the I-band, spectral
classification is efficient. One of the objects studied
(2MASSW J1523+3014) is the coolest L-dwarf discovered so far by the
2-Micron All-Sky Survey (2MASS), but its spectrum is still 
significantly different from the methane-dominated
objects such as Gl229B or SDSS 1624+0029.

\end{abstract}


\section{Introduction}
After eluding undisputed detection for many years, numerous brown dwarfs
-- objects with 
sub-stellar mass -- are now known. While some candidates were discovered
in small-scale
surveys of
young nearby clusters, such as the Pleiades and Hyades, or as companions
to low-mass stars, 
the biggest breakthrough has come as a result of large-scale surveys such
as the Deep
Near-Infrared Sky (DENIS) survey (Delfosse \et 1997), 2MASS, the 2-Micron
All-Sky Survey
(Skrutskie \et 1997, Kirkpatrick \et 1999) and the Sloan Digital Sky
Survey (SDSS) 
(Strauss \et 1999). 
Recently, using optical (CCD) spectroscopy, Kirkpatrick \et (1999) have
defined a new spectral class, L-dwarfs, in which the metallic oxides (such
as TiO and VO) found 
in M stars lose their dominance to metallic hydrides (such as FeH and
CrH). The temperature
range for the L class is given by Kirkpatrick \et (2000) from about 
2000 K for L0 to about 1250 K for L8, whereas Martin \et (1999) suggest a 
range from 2200-1600 K. Depending on age and model
calculations, Kirkpatrick \et 
(1999)
argue that at least one third of the L-dwarf objects must be brown dwarfs
and perhaps all are.

Spectral classification is based on spectroscopy between 6500 and 10000
\AA~ at a resolution of 9\AA~ (R $\sim$1000). 
While use of this spectral region provides many important spectral
diagnostics, it suffers from the 
fact that L- and T-dwarfs are extremely faint at these wavelengths and
therefore long exposures
on very 
large aperture telescopes are required to obtain spectra with good
signal-to-noise ratios. Typical
I-band magnitudes are about 19 or fainter (\eg~ GD165B), but a gain of 3-6
magnitudes can be
obtained in going to the near infrared.

Using earlier generations of infrared instruments, previous observations
of individual brown dwarf
candidates have yielded a typical resolving power of about R=500-1000; see
the
observations of Geballe \et (1996), Ruiz, Leggett and Allard (1997), Tinney,
Delfosse and Forveille (1997), Kirkpatrick \et (1999) and Strauss \et
(1999). These pioneering
efforts were accomplished with instruments using an earlier generation of
IR detector arrays,
with at most 256 x 256 pixels. This resolution is sufficient to reveal the
major
differences that set apart the L-dwarfs and T-dwarfs from warmer stars,
\eg, the presence of deep steam bands and strong
methane bands in the
L- and T-dwarfs respectively. Kirkpatrick \et (1993) modeled a spectral
sequence of M-dwarfs
using spectroscopy from 0.6 - 1.5 microns and identified the major bands
and atomic features.
Jones \et (1996) performed a similar analysis from 1.16 - 1.22 microns
with a sample of M dwarfs
which also included GD165B. An excellent review of model atmospheres of
very low mass stars
and brown dwarfs is given by Allard \et (1997).

In this paper we report observations using
NIRSPEC, a new cryogenic infrared spectrograph on the Keck II telescope
employing
a 1024 x 1024 InSb array.  A consistent set of J-band spectra with
R$\sim$2500 is presented
which, for the first time, allows a detailed comparison of the
near-infrared features of the spectral
sequence from early L-dwarfs to T-dwarfs. Our targets were
selected from the
list of L-dwarfs published by Kirkpatrick \et (1999) and supplemented with
new sources
discovered more recently by the 2MASS (Kirkpatrick \et, 2000). 
One of these objects is reported as being the closest
known L-dwarf to date and another is likely the coolest L-dwarf discovered
thus far.

\section{Observations}

Table 1 lists the objects observed and provides a summary of their
photometric properties and 
spectral classification based on the far-optical spectroscopy 
by Kirkpatrick \et (1999). As part of the
``first light" scientific commissioning of the NIRSPEC 
spectrograph at the W. M. Keck Observatory on Mauna Kea, Hawaii,
near-infrared spectra of this 
sample were obtained. Kelu-1 (Ruiz \et, 1997), was observed on April 29,
1999, but all of the other sources were observed on June 2, 1999. Since
detailed descriptions of the
design and performance of NIRSPEC are given elsewhere, (McLean \et 1998,
McLean \et 2000),
only a short summary is included here.
Briefly, this cryogenic instrument is the
world's first facility-class infrared spectrograph employing the 
state-of-the-art 1024 x 1024 InSb array. For the highest spectral resolution 
work, a cross-dispersed echelle grating is used  which yields R=25000 for 
a 0.43\arcs~ wide slit, corresponding to 3-pixels along the dispersion
direction. A much lower resolution mode can be obtained simply by replacing 
the echelle grating with a flat mirror and 
using the cross-dispersion grating alone. The spectral resolution in this
mode is R $\sim$~ 2500 for
a 2-pixel wide slit (corresponding to 0.38\arcs~ in this case). 

For the present study, the lower resolution mode was selected for speed
and efficiency. The goal was to obtain
a spectral sequence of L-dwarfs with good signal-to-noise ratios from about 
1-2.5 \mic. Only the J-band results, covering
the interesting range from 1.135 -- 1.362 \mic~ are discussed
in this short note.

As shown in Table 1, the J magnitudes of the sample range from 12.8 to
16.3. All objects
received the same total exposure time of 600 seconds. The observing
strategy employed was to
obtain a 300 s integration at each of two positions along the entrance
slit separated by about
20\arcs~, referred to as a nodded pair. Seeing conditions were generally
very good for these
measurements (0.3\arcs -- 0.5\arcs) and a slit width of 0.38\arcs~ was
used in all cases. 
To calibrate for
absorption due to the Earth's atmosphere, stars of spectral type A0 V to
A2 V were observed as
close to the same airmass as possible (typically within 0.05 airmasses,
except for 2MASSW
J1632+1904 for which the difference was 0.28) and also close in time. The
J-band is 
sensitive to atmospheric extinction due mainly to water vapor absorption.
A-type stars are
essentially featureless in this region except for the Paschen Beta line at
1.2816\mic, which can be 
interpolated out. Immediately after the observation of each source,
both neon and argon arc lamp spectra were obtained for wavelength
calibration, and a white-light
spectrum was recorded for flat-fielding.

Reduction of the data followed the steps set out below. The first
requirement is to place the raw
data on a uniform grid of wavelength and position along the slit. Using
custom software
developed by one of us (James Larkin) the spatial distortions were
corrected first. A spatial map
was formed by using the sum of the nodded pair of standard star spectra
with the assumption that
the pair of spectra must be exactly a fixed number of pixels apart. Next,
the arc lamp spectra were
used to construct a spectral map to warp the raw data onto a uniform
wavelength scale using a
second order polynomial fit. Next, the A-type calibration star 
was reduced by forming the difference image, warping it with the spatial
and spectral mapping
routines, dividing by the normalized flat field lamp, shifting and
co-adding the pair of spectra at
the two slit positions and then extracting the resultant spectrum.
Division with a
blackbody spectrum for the temperature corresponding to the star's
published spectral class
completed the reduction of the standard star. Finally, the Paschen Beta
absorption line 
at 1.2816\mic~ was removed by interpolation from the reduced spectra
before it was used for
division into the corresponding object spectra.

Similar steps were applied to the raw data frames of the target sources.
After rectification and
flat-field correction, each spectrum was extracted and divided by the
fully-reduced spectrum of
its associated calibration star. Finally, the nodded pair of reduced
spectra were
shifted, co-added together and extracted to give the resultant calibrated
spectrum of each source.
The results are shown in Figure 1. Note that this
entire set of new infrared
spectra for a sample of seven optically-faint, low-mass objects represents
a total of only 70
minutes of on-source observing time, comparable to the exposure time {\it
per} object needed
with other instruments or at shorter wavelengths.

\section{Results}
Clearly, the strongest atomic line transitions in this wavelength region
are the pair of neutral
potassium (K I) lines at 1.1690, 1.1770\mic~ and 1.2432, 1.2522\mic~
respectively. The first pair
correspond to the multiplet designation 4p $^{2}$P$^{o}$ - 3d $^{2}$D, and
the second pair are
from the 4p $^{2}$P$^{o}$ - 5s $^{2}$S multiplet. The dominant molecular
species in the
L-dwarfs in this band are H$_{2}$O and iron hydride (FeH), with methane
(CH$_{4}$)
appearing in the T-dwarf. The strongest FeH bands are expected at 1.194,
1.21 and 1.237\mic~
approximately. A pair of sodium lines, the 3p $^{2}$P$^{o}$ - 4s $^{2}$S
multiplet, can just be detected buried in the water absorption at 1.138
and 1.141\mic~ and we
report the detection of a weak rubidium line at 1.3233\mic~ (5p
$^{2}$P$^{o}$ - 6s $^{2}$S) as
well as the cesium line at 1.3588\mic~ (6p $^{2}$P$^{o}$ - 7s $^{2}$S) in
the spectrum of
2MASSW J1507. Many other features are evident however, and even the
smallest spectral
structures are real and above the noise level. Distinctive
patterns of lines around 1.25-1.33\mic~ repeat from object to object among
the earlier spectral
types but fade out in the later spectral classes. Comparison of the region from
1.16--1.22\mic~ in our L-dwarf spectra with the same region studied in
M-dwarfs by Jones \et
1996 (see their Fig. 2 for Gl 406) using CGS4 on UKIRT, reveals excellent 
detailed agreement.

Although the spectra are still heavily blended even at R$\sim$2500, we
have extracted equivalent 
widths and full widths at the base of each line for the four K I
transitions. We have also
constructed an index to measure the strength of the H$_{2}$O absorption
using the ratio of the
flux at 1.33\mic~ to that at 1.27\mic. The results are summarized in Table
2. The equivalent
width of the K I lines changes very slowly from an average of about 7 \AA~
across the L-dwarfs
to about 12 \AA~ in the T-dwarf SDSS1624, although the line depth decreases
significantly. This behavior
is due to line broadening. The full width at the base of these lines
increases from about
40\AA~ in the L-dwarfs to over 80 \AA~ in the T-dwarf. The water index
slowly strengthens
from about 0.6 to 0.4 through the L-dwarfs as the absorption at 1.33\mic~
becomes deeper and
then drops markedly to about 0.04 in the T-dwarf.

\section{Discussion}

The variation in the J-band spectra of this sample of objects is quite
remarkable and it is relatively
easy to place the objects in a temperature sequence. The water band
strengthens as the
temperature decreases. FeH weakens and then disappears, the K I lines
weaken and broaden and
the continuum around 1.15\mic~ slowly drops relative to the continuum at
1.26\mic. As expected,
the DENIS L5 source and the 2MASS L5 object exhibit almost identical
spectral characteristics. 
By ordering the spectra according to the classifications given by
Kirkpatrick \et (1999, 2000),
with the earliest spectral type (L2) at the top, the following trends are
apparent in the J-band spectra.\\ 
{\it L2 (Kelu-1)}: strong K I and FeH lines are superimposed on a larger
depression across the
region, which is perhaps the result of residual oxide (either TiO and/or
VO) absorption;  VO is expected around 1.19\mic.\\
{\it L4 (GD165B)}: any residual oxide absorption has gone, effectively
raising the continuum to
produce a flatter spectrum, and making the K I and FeH features appear
stronger although they
are expected to decrease with decreasing temperature. The water absorption
(steam) band at
1.30\mic~ is increasing in strength. Numerous small features from
1.25--1.30\mic~ closely match
those in Kelu-1.\\
{\it L5 (2MASSW J1507-1627)}: all features present in the L4 class remain.
The K I and FeH
features are very slightly weaker, while the water band at 1.30 microns is
deeper than before
and there is a slight slope of the continuum towards the blue end.\\
{\it L5 (DENIS-P J1228-1547)}: this spectrum is almost the same as the
previous one,
confirming that they are indeed the same spectral class.\\
{\it L8 (2MASSW J1623+1904)}: at L8, the FeH features have disappeared and
the depth of the
K I lines are significantly weaker but there is evidence of broadening in
their wings. There is a
slight downward slope of the continuum towards the blue. The steam band is
relatively
stronger.\\
{\it L8/9 (2MASSW J1523+3014)}: Very similar to the previous L8, but the K
I lines appear
slightly broader and the water band is slightly deeper. The slope to the
blue is a little stronger than
in L8. Consequently, this object may be cooler than 2MASSW J1623+1904 as
its designation
suggests, but the difference is small.\\
{\it T (SDSS 1624+0029)}: A dramatic slope towards the blue appears, due
to the
onset of methane absorption in this wavelength region, and there is also a
slope or ``break"
towards the red from about 1.26--1.31\mic~ before a deep water band sets
in. The K I lines are
still present but are now very broad. 

As an illustration of the density of molecular features and the problem of line
blending, Figure 2 shows a model spectrum kindly provided by Peter Hauschildt
(private communication). This sample spectrum is based on a model atmosphere 
code (AMES-Dusty, Allard \& Hauschildt. in prep), 
with a self-consistent treatment for dust formation. The treatment of dust is 
complicated however, and theorists have yet to agree on the best approach. 
The parameters of this model are 
solar metallicity, log(g)=4.5 and T$_{eff}$=2000 K and the model spectrum 
was smoothed from an original resolution of R=50,000. Qualitatively, 
the agreement with the NIRSPEC spectra is very good.

Another useful framework for understanding these spectra is the molecular
equilibrium
calculations by Burrows and Sharp (1999). As their analysis shows, the
main absorbers
characteristic of M stars (\eg TiO and VO) decline rapidly in importance
with decreasing effective
temperature. These molecules are expected to condense onto dust grains;
TiO for
instance forms perovskite (CaTiO$_{3}$). The abundance of gaseous TiO
begins to decrease
around 2400 K and similarly, VO will become depleted near 1800 K. For
iron, the first
condensate to form is the metal itself, at about 2200 K, which can then
form droplets and rain out
of the atmosphere. We have carefully compared our spectra to the solar
atlas and cannot make
any conclusive identifications with iron lines, or any other metal 
lines (such as Mn and Al) among the dense forest of H$_{2}$O transitions.
Interestingly, Jones \et 1996 noted 
the presence of Fe in earlier spectral types, such as the M6 dwarf GL406,
at a comparable resolution. A significant amount of iron may have rained 
out. 

Since they are less refractory and survive in monatomic form for a greater
temperature range, the
neutral alkali metals (Na, K, Rb, Cs) are expected to remain after the
true metals become depleted. 
In effect, as the temperature falls the atmospheres of cool
sub-stellar objects become
more transparent. The column density of potassium and sodium, for
instance, is
expected to increase to the point where the wings of the absorption lines
become damped. This
result explains the strength, broadening and temperature dependence of the
K I lines seen in our
spectra. According to Burrows and Sharp (1999), sodium and potassium
should become depleted
around 1500 - 1200 K, with sodium disappearing first and potassium forming
into KCl below about
1200 K. If there is settling of refractory species however, at higher, deeper 
temperatures, then both atomic sodium and potassium are expected to 
persist to lower 
temperatures, at which point they should form their sulfides, not 
chlorides (see Burrows, Marley and Sharp 1999, and Lodders 1999). 
Figure 1 shows that the very strong K I lines persist, albeit 
with broad wings, well into the T-dwarf temperature range.

Some features apparent in the new data are not yet explained by the
existing models. For example,
a broad, relatively strong feature is seen in our spectra at 1.22 \mic.
This feature remains through L5, but is gone in the L8 spectra. Although 
this is the same pattern as followed by FeH, this broad feature does not 
appear in the opacity plot of FeH kindly supplied by Adam Burrows (private 
communication), nor in the model spectrum provided by Peter Hauschildt. Finally, 
our results imply that any L- or
T- dwarf object meeting the discovery parameters of the 2MASS and/or the
SDSS can be
observed spectroscopically with NIRSPEC on Keck at medium to high spectral
resolution. The
near-infrared region from 1.13 - 1.36 \mic~ is quite rich in spectral
features, most of which
appear to be unresolved blends of molecular species, namely H$_{2}$O and
FeH in the L-dwarfs
and CH$_{4}$ in the T-dwarfs. Evidently, even higher spectral resolution 
would help to constrain the models.\\

\acknowledgements
It is a pleasure to acknowledge the hard work of past and present members
of the NIRSPEC
instrument team at UCLA:  Maryanne Anglionto, Odvar Bendiksen, George
Brims, Leah
Buchholz, John Canfield, Kim Chim, Jonah Hare, Fred Lacayanga, Samuel B.
Larson, Tim Liu,
Nick Magnone, Gunnar Skulason, Michael Spencer, Jason Weiss, Woon Wong.  
In addition we thank director Fred Chaffee, CARA instrument specialist
Thomas A. Bida, 
and the Observing Assistants at Keck observaory,  Joel Aycock, Gary
Puniwai, Charles
Sorenson, Ron Quick, and Wayne Wack, for their support. We are pleased to
acknowledge the
International Gemini Telescopes Project for the InSb detector used in
these measurements.
Finally, we gratefully acknowledge Adam Burrows and Peter Hauschildt for
very helpful
information and advice about the current model atmospheres of low-mass
stars and sub-stellar
objects.\\

\clearpage
\begin{deluxetable}{lccccccc}
\small
\tablewidth{0pt}
\tablecaption{\bf Observed Objects and Summary of their Photometric
/Spectral Properties}

\tablehead{\colhead{Object} & \colhead{RA(2000)} & \colhead{Dec(2000)} &
\colhead{J} &
\colhead{H} & \colhead{K$_{s}$} & \colhead{Sp. Type} & \colhead{Ref.}}

\startdata
DENIS-P J1228.2-1547  &  12 28 13.9  &  -15 47 11.7  &  14.38  &  13.37  &
 12.81  &  L5   &
1\nl
Kelu-1                &  13 05 40.2  &  -25 41 06    &  13.38  &  12.41  &
 11.81  &  L2   & 1\nl
GD165B                &  14 24 40.0  &  +09 17 24    &  15.80  &  14.78  & 
14.17\tablenotemark{a}
&  L4   & 1\nl
2MASSW J1507-1627      &  15 07 47.7  &  -16 27 39    &  12.82  &  11.89
&  11.30  &  L5  
&
2\nl
2MASSW J1523+3014      &  15 23 22.6  &  +30 14 56    &  16.32  &  15.00
&  14.24  &  L8/9
&
2\nl
2MASSW J1632+1904     &  16 32 29.4  &  +19 04 41    &  15.86  &  14.59  &
 13.98  &  L8  
&
1\nl                     
SDSS 1624+0029        &  16 24 14.4  &  +00 29 15.6  &  15.53  &  15.57  & 
15.70\tablenotemark{a} &  T    & 3\nl
\enddata
\label{t:irlog}

\tablenotetext{a}{This is K, not Ks.}
\tablenotetext{ }{1.Kirkpatrick, J.D., Reid, I.N., Liebert, J., Cutri,
R.M., Nelson, B., Beichman,
C.A., Dahn, C.C., 
Monet, D.G., Gizis, J.E., and Skrutskie, M.F. 1999, ApJ, 519, 802 }
\tablenotetext{ }{2.Kirkpatrick, J.D., \et, 2000, ApJ, in prep}
\tablenotetext{ }{3.Strauss, M. A., \et, 1999, ApJ, 522, L61 }

\end{deluxetable}

\clearpage
\begin{deluxetable}{lcccccc}
\small
\tablewidth{0pt}
\tablecaption{\bf Equivalent Widths (\AA) of the K I lines and the
Strength of the H$_{2}$O
Band at 1.33 \mic}

\tablehead{\colhead{Object} &\colhead{Sp.Type} & \colhead{1.169\mic} &
\colhead{1.177\mic} & \colhead{1.244\mic} &
\colhead{1.253\mic} & \colhead{1.33\mic/1.27\mic}}

\startdata
Kelu-1                           & L2   & 6.18  &   6.86  & 6.64  &  6.32
& 0.55  \nl
GD165B                        & L4   & 6.79  &   9.34  & 8.64  &  7.52  &
0.56  \nl
DENIS-P J1228.2-1547& L5   & 7.26  &   9.94  & 7.88  &  7.06  & 0.52  \nl
2MASSW J1507-1627  & L5   & 8.98  & 10.82  & 7.68  &  7.84  & 0.46  \nl
2MASSW J1632+1904 & L8   & 7.75  &   8.35  & 6.02  &  7.33  & 0.38  \nl  
2MASSW J1523+3014 & L8/9& 8.83  &   9.18  & 6.17  & 6.56   & 0.41  \nl
SDSS 1624+0029          & T     & 8.93  & 14.75  & 9.09  & 11.38 & 0.04  \nl
\enddata
\label{t:irlog}

\end{deluxetable}

\clearpage

\begin{figure}[t]
\caption{NIRSPEC spectra with a resolving power of R=2,500 (5 \AA) and a
dispersion of 2.5 \AA~ per pixel from 1.135 to 1.357 \mic~ for a sample of
6 L-dwarfs and one
T-dwarf. Each spectrum has been normalized to unity using the average flux
near 1.28 \mic~ and
then displaced by 0.6 units along the y-axis for clarity of presentation.
For the fainter
sources, the first few and last few data points are too noisy to plot.
Prominent features are
identified, but much of the structure is due to blending of molecular
transitions.}
\end{figure}

\begin{figure}[t]
\caption{An example of a model spectrum from Hauschildt \et with a final (smoothed)
resolution of R=2,000 (6
\AA) from 1.135 to 1.357 \mic~ for qualitative comparison with the NIRSPEC data. 
The model parameters are log(g)=4.5, T$_{eff}$=2000 K and solar metallicity. Lines identified in the model spectrum are not necessarily the same as those seen in the NIRSPEC observations (see text).}
\end{figure}


\begin{references}

Allard, F., Hauschildt, P.H., Alexander, D.R. and Starrfield, S. 1997,
Annu. Rev. Astron.
Astrophys., 35, 137.\\
Burrows, A. and Sharp, C.M. 1999, ApJ, 512, 843 \\
Burrows, A., Marley, M.S. \& Sharp, C.M. 1999, in press\\
Delfosse, X., \et 1997, A \& A, 327, L25 \\
Geballe, T.R., Kulkarni, S.R., Woodward, C.E. and Sloan, G.C. 1996, ApJ,
467, L101\\
Jones, H.R. A., Longmore, A. J., Allard, F., and Hauschildt, P. H. 1996,
MNRAS, 280, 77\\
Kirkpatrick, J.D., Kelly, D.M., Rieke, G.H., Liebert, J., Allard, F.,
Wehrse, R. 1993, ApJ, 402,
643 \\
Kirkpatrick, J.D., Reid, I.N., Gizis, J.E., Burgasser, A.J., Liebert, J., 
Monet, D.G., Dahn, C.C., and Nelson, B. 2000, ApJ, in prep.
Kirkpatrick, J.D., Reid, I.N., Liebert, J., Cutri, R.M., Nelson, B.,
Beichman, C.A., Dahn, C.C., 
Monet, D.G., Gizis, J.E., and Skrutskie, M.F. 1999, ApJ, 519, 802 \\
Lodders, K. 1999, ApJ, 519, 793\\
Martin, E. L., Delfosse X., Basri, G., Goldman, B., Forveille, T., \& 
Zapatero-Osorio, M. R. 1999, AJ, 118, 2466\\
McLean, I.S., \et 1998, SPIE, 3354, 566\\
McLean, I.S., \et 2000, PASP, in preparation \\
Ruiz, M.T., Leggett, S.K. \& Allard, F. 1997, ApJ, 491, L107 \\
Skrutskie, M. F., \et 1997, in The Impact of Large-Scale Near-IR Sky
Surveys, ed. F. Garzon (Dordrecht: Kluwer), 25\\
Strauss, M. A., \et, 1999, ApJ, 522, L61\\
Tinney, C.G. \et, 1993, AJ, 105, 1045.\\
Tinney, C.G., Delfosse, X., and Forveille, T. 1997, ApJ, 490, L95.\\
\end{references}
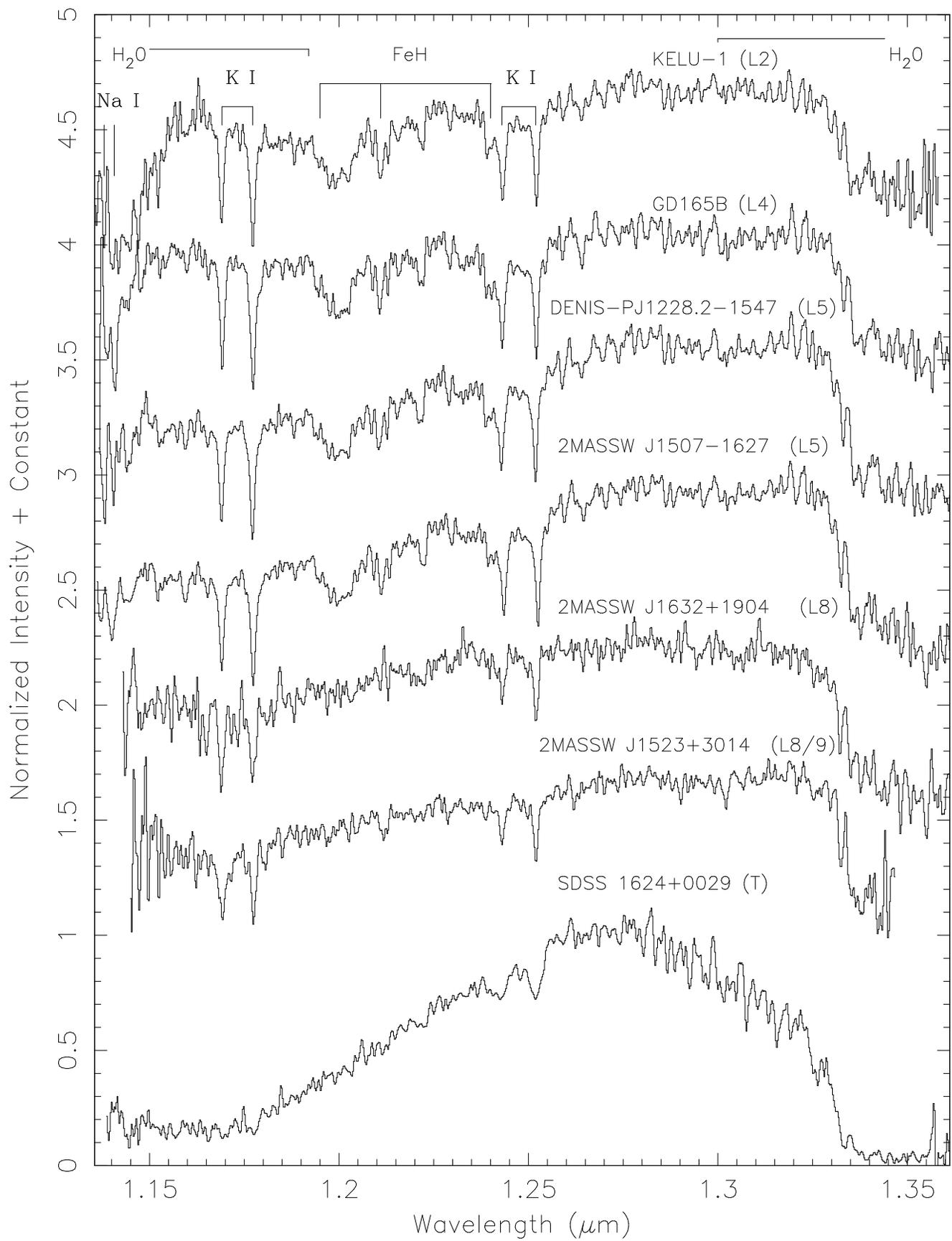
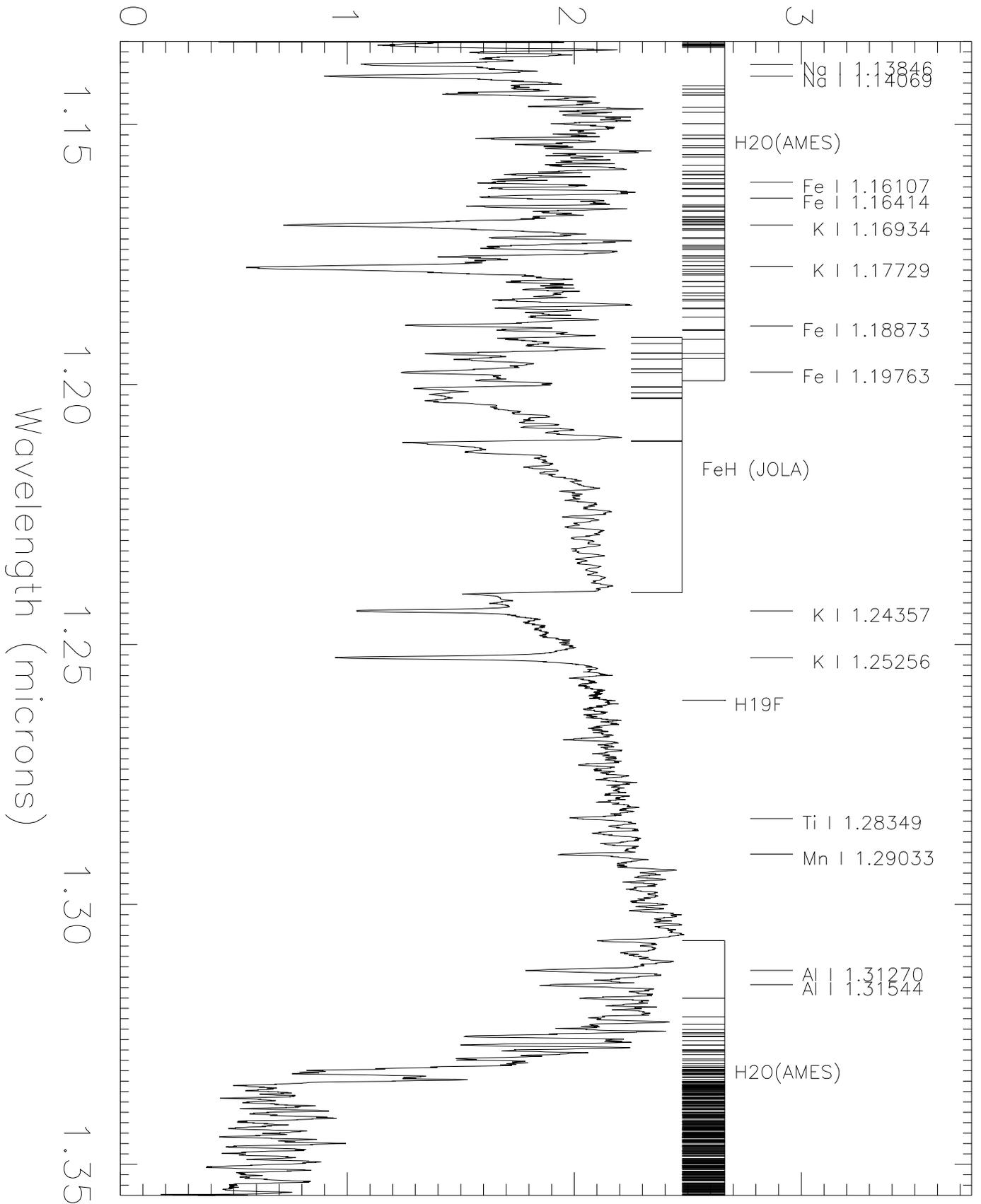
\end{document}